\documentclass[doublecol]{epl2}
\usepackage{amsmath}
\usepackage{amssymb}
\usepackage{amsfonts}
\usepackage{subfigure}
\usepackage{float}
\usepackage{url
}

\newcommand{\J}{ \ab{J_0} }

\newcommand{\ka}{\kappa_\ab{A}}
\newcommand{\kb}{\kappa_\ab{B}}
\newcommand{\kab}{\kappa_\ab{AB}}

\title{Coherently controlled entanglement generation in a binary Bose-Einstein condensate}
\shorttitle{Entanglement generation in a binary BEC}
\author{Niklas Teichmann\inst{1}\thanks{\email{teichmann@theorie.physik.uni-oldenburg.de}}  \and Christoph Weiss\inst{2}\thanks{\email{christoph.weiss@lkb.ens.fr}}}
\shortauthor{N. Teichmann \etal}

\institute{
  \inst{1} Institut f\"ur Physik - Carl von Ossietzky Universit\"at - D-26111 Oldenburg - Germany\\
  \inst{2} Laboratoire Kastler Brossel - \'Ecole normale sup\'erieure - 24 rue Lhomond -
  		F-75231 Paris Cedex 05 - France
}

\abstract{
	Considering a two-component Bose-Einstein condensate in a
	double-well potential,
	a method to generate a Bell state
	consisting of two spatially separated condensates is suggested.
	For repulsive interactions, the required tunnelling control
	is achieved numerically by varying the amplitude of a
	sinusoidal	potential difference between the wells.
	Both numerical and analytical calculations reveal the
	emergence of a highly entangled mesoscopic state.
}

\begin{document}

\maketitle

	In a recent ground-breaking experiment self-trapping
	was observed for small Bose-Einstein condensates (BECs)
	in a double-well potential~\cite{Oberthaler2005}.
	A major experimental goal in this system is the generation of mesoscopic
	entangled number states
	like ``Schr\"odinger-cat'' states \cite{Oberthaler2005}.
	Unfortunately, in a double well the ground state of a condensate with repulsive interaction
	is not the desired ``Schr\"odinger-cat'' state (e.g.\ ref.~\cite{HolthausStenholm2001}).
	On the other hand using attractive condensates leads to the problem of
	instability (e.g.\ ref.~\cite{DoddEdwards1996}).
	However, a two-component repulsive
	condensate in a double well is investigated in this letter, because
	it offers the possibility to increase the entanglement of the ground state
	drastically.

	To control the entanglement generation, we employ
	a method recently suggested to induce the phase transition between a superfluid and a Mott
	insulator~\cite{GreinerMandelEsslinger2002}
	by applying time-periodic potential differences~\cite{EckardtWeissHolthaus2005,CreffieldMonteiro2006}.
	While on the single particle level the effects predicted for periodically
	driven systems~\cite{GrossmannDittrichJung1991,Holthaus1992} have not yet
	been realised experimentally, the coherent control by changing the amplitude 
	of periodic force fields might be easier to achieve for
	BECs rather than for single particles.

	Suggestions to produce mesoscopic entangled states,
	involve coherent scattering of far detuned
	light fields~\cite{RuostekoskiCollettGraham1998}; several groups suggest to produce
	mesoscopic entanglement with condensates
	by manipulating the interaction between the particles, or by controlling the
	dynamics of the	system~\cite{CiracLewensteinMolmer1998,SorensenDuanCirac2001,
	DunninghamBurnett2001, MicheluJakschCirac2003,MahmudPerryReihardt1998,Jinasundera_Weiss}. 
	While the experimental realisation of such mesoscopic entangled states is still a challenge 
	of current fundamental research, possible applications include quantum computing or quantum 
	information processing.

	In this letter we study a double-well potential similar
	to the experiment~\cite{Oberthaler2005}, but with
	a two-component condensate (for various aspects investigated for two-component BECs see
	e.g.\ refs.~\cite{MyattBurtGhrist1997,Stamper-SturnAndrewsChikkatur1998,MaddaloniModugnoFort2000,GoldsteinMeystre1997}).
	To motivate the mechanism by which we control the entanglement generation, we
	start with the two-particle case, because it exhibits all crucial
	features, before we extend both the numerical and analytical calculations to $N$ particles.

\section{The model}

	We study a symmetric double-well potential filled with Bose particles
	at very low temperature.
	Adopting the common two-mode approximation, assuming only on-site interactions
	and denoting the tunnelling
	splitting between the two lowest single particle energy states with $\hbar \Omega$,
	we use a well established model describing the dynamics
	of a one-species BEC in double-well potential
	\cite{MilburnCorneyWright1997}. This model
	gives a good qualitative description of recent experiments \cite{Oberthaler2005}.
	Extending the model to a binary condensate consisting of $N$ particles of
	species A and $N$ particles of species B leads to the Hamiltonian
\begin{eqnarray}
	\label{N-Hamiltonian}
 \hat{H} &=& -\frac{\hbar \Omega}{2}(\hat{a}_1 \hat{a}^{\dagger}_2 + \hat{a}^\dagger_1 \hat{a}_2)
			+ \hbar \ka (\hat{a}^{\dagger}_1 \hat{a}^{\dagger}_1 \hat{a}_1 \hat{a}_1
			+ \hat{a}^{\dagger}_2 \hat{a}^{\dagger}_2 \hat{a}_2 \hat{a}_2 )
			\nonumber\\
		   &&-\frac{\hbar \Omega}{2}(\hat{b}_1 \hat{b}^{\dagger}_2 + \hat{b}^\dagger_1 \hat{b}_2)
			+ \hbar \kb (\hat{b}^{\dagger}_1 \hat{b}^{\dagger}_1 \hat{b}_1 \hat{b}_1
			+ \hat{b}^{\dagger}_2 \hat{b}^{\dagger}_2 \hat{b}_2 \hat{b}_2 )
			\nonumber\\
			&&+ 2\hbar \kab(\hat{a}^{\dagger}_1 \hat{a}_1 \hat{b}^{\dagger}_1 \hat{b}_1
			+  \hat{a}^{\dagger}_2 \hat{a}_2 \hat{b}^{\dagger}_2 \hat{b}_2)
			\nonumber\\
			&&+ \hbar \mu f(t) (\hat{a}^{\dagger}_1 \hat{a}_1 - \hat{a}^{\dagger}_2 \hat{a}_2
			+ \hat{b}^{\dagger}_1 \hat{b}_1 - \hat{b}^{\dagger}_2 \hat{b}_2	)\;,
\end{eqnarray}
	where $\hat{a}_i^{(\dagger)}$ and $\hat{b}_i^{(\dagger)}$ are the annihilation (creation)
	operators for a boson of species A or B in the $i$th well ($i$ = 1,2) and
	$\hbar \mu f(t)$ specifies an externally applied potential difference between the
	two wells.
	The relation between the interaction parameters $\ka$, $\kb$ and $\kab$
	and the experimental situation can be found e.g.\ in ref.~\cite{HolthausStenholm2001}.
	We use the Fock-basis
	$| i, N-i \rangle_\ab{A} | j, N-j \rangle_\ab{B} \equiv
	| i, j \rangle$, where $i$/$j$ are the numbers of A/B particles
	occupying the first well and $(N-i)$/$(N-j)$ the number of A/B particles
	occupying the second well.

	In these notations, the 
\emph{Bell state} reads:
	\begin{eqnarray}
		\psi_\ab{Bell} &=&
		\frac{1}{\sqrt{2}} \big( | N, 0 \rangle + |0, N\rangle \big)
		\nonumber\\
		&\equiv& \frac{1}{\sqrt{2}} \big(|N, 0\rangle_\ab{A}|0,N \rangle_\ab{B}
		+ |0,N \rangle_\ab{A}|N,0 \rangle_\ab{B} \big)\;.
		\label{Bell}
	\end{eqnarray}
	In this state, the two condensates are maximally entangled with each other in the sense that each 
        condensate is in a superposition of being in both wells and if one condensate is measured in one well, 
        the other condensate will be in the other well.

	Similar spatially entangled states are known from the EPR-paradox and
	are of special interest
	in the field of quantum cryptography, where especially entangled photons
	are used \cite{JenneweinSimonWeihs2000,Nielsen_Chuang}.
	Looking at the $N$-particle dynamics, we have to take $(N+1)^2$ states into account.
	If we exemplarily study the case of only two distinguishable
	particles A and B without driving
	the well ($\mu = 0$), the system can be solved analytically. While we assume short range interaction between 
	the two particles typical for cold atoms, 
	the resulting entanglement is similar to the case of two electrons in a double
        quantum dot~\cite{ZhangEtAl02}. 
	Here, we use the basis
	$\{|1,1 \rangle, |1,0 \rangle, |0,1 \rangle, |0,0 \rangle \}$ to obtain the system's
	Hamiltonian as a $4 \times 4$ matrix
	\begin{equation}
		\hat{H}_2 = \hbar\Omega
		\begin{pmatrix} u & - 1 /2   & - 1 /2 & 0 \\
     - 1 /2 & 0 & 0 & - 1 /2\\
      - 1 /2 & 0 & 0 & - 1 /2\\
      0 & - 1 /2 & - 1 /2 & u \end{pmatrix}\; ,
	\end{equation}
	where $u = \kab/\Omega$ is the interaction parameter.
	When repulsive interaction is assumed ($u > 0$),
	the ground state and first excited state of the system are given by
	\begin{equation}
		\vect{y}_0 = \mathcal{N}
		\begin{pmatrix} 1 \\ \frac{u}{2}+\frac{1}{2}\sqrt{u^2 + 4} \\
    \frac{u}{2}+\frac{1}{2}\sqrt{u^2 + 4} \\ 1 \end{pmatrix}
	\;;\quad
	\vect{y}_1 = \frac1{\sqrt{2}}\cdot\begin{pmatrix} 0 \\ -1 \\
    1 \\ 0 \end{pmatrix}\
	\end{equation}
	with norm $\mathcal{N} = \left(4+u^2+u\sqrt{u^2+4}\right)^{-1/2}$
	and corresponding energies $E_0=\hbar \Omega(u/2-\sqrt{u^2+4}/2)$ and $E_1=0$.
	In the case of a very large interaction parameter the ground state
	approaches the \emph{Bell state} (eq.~(\ref{Bell}) with $N=1$),
	\begin{equation}
		\vect{y}_0
		\xrightarrow{u \to \infty}
		\frac{1}{\sqrt{2}}	\begin{pmatrix}
			0 & 1 & 1 & 0
		\end{pmatrix}^t
		=\frac{1}{\sqrt{2}} \big( |0,1 \rangle  + |1,0 \rangle \big)\;,
	\end{equation}
	and becomes a maximally entangled state.
	On the other hand with no interaction ($u=0$) the ground state is given by
	$1/2 \cdot \begin{pmatrix} 1 & 1 & 1 & 1 \end{pmatrix}^t
	=1/2\cdot \left( |1, 1 \rangle + |0, 1 \rangle + |1, 0 \rangle + |0, 0 \rangle \right) $,
	which can be written as a product state
	$\vect{y_0}(u=0)=1/2\left( |1,0 \rangle_\ab{A} + |0,1 \rangle_\ab{A} \right) \cdot
	\left( |1,0 \rangle_\ab{B} + |0,1 \rangle_\ab{B} \right) $ and therefore is not
	entangled at all.
	Furthermore, for very large interaction parameter $u$
	the ground state is quasi-degenerate.
	This is also true for the $N$-particle system with the states
	$(|N,0  \rangle \pm |0, N \rangle)/\sqrt{2}$. Because of this
	quasi-degeneracy one cannot generate a \emph{Bell state}
	by just generating a binary-BEC in a very deep double-well potential; the system could be in 
	any superposition of those two states. If, however, one starts in the ground state of a not so deep well, 
        the Hamiltonian does not allow transitions between
	symmetric and antisymmetric states (cf.\ ref.~\cite{Haroche}). Thus, thermal excitations to the
	(antisymmetric) first excited state can be discarded.

	The double-well potential is now
	modulated periodically in time by the force $f$
	indicated in eq.~(\ref{N-Hamiltonian}). As the new Hamiltonian
	is periodic in time, Floquet theory can be applied.
	As shown in refs.~\cite{Tsukada1999,EckardtJinasunderaWeiss2005}, for sufficiently high
	frequencies ($\omega/\Omega \gg 1$) the system behaves
	like an undriven system with a rescaled
	effective tunnelling frequency
	$	\Omega_\ab{eff} = \Omega \cdot \J\left(2\mu/\omega \right)$
	or equivalently with the effective interaction parameter
	\begin{equation}
		u \to u_\ab{eff} = \frac{u}{\J\left(\frac{2\mu}{\omega}\right)}\;,
		\label{ueff}
	\end{equation}
	where $\J$ denotes the ordinary Bessel function of order zero.
	The system is coherently controlled by driving it with linearly
	in time increasing parameter $2\mu/\omega$ (in this case $2\mu/\omega \simeq
  	2.405 \cdot \tau/\tau_\ab{max}$, where $\tau = \Omega \cdot t$
	is a dimensionless time variable). At the time $\tau_\ab{max}$
	the parameter reaches the first zero of $\J$;
	the effective interaction becomes very large. Similar effects could be achieved 
	without driving by
	adiabatically increasing the depth of the wells.
	Compared with this coherent control mechanism, the feed-back loops
	of optimal control theory~\cite{AssionBaumert1998}
	would allow
	to further optimise entanglement generation for given experimental parameters.

\begin{figure}
	\includegraphics[scale=0.28,angle = -90]{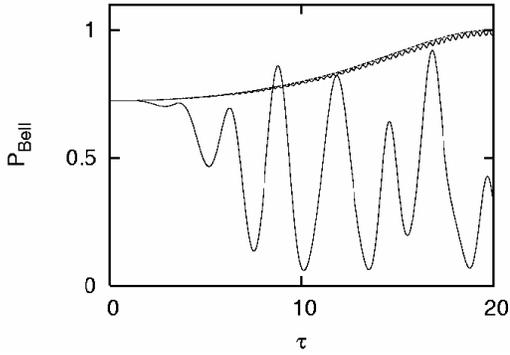}

	\caption{Fidelity (eq.~(\ref{fidelity}), probability of the system
			to be in the \emph{Bell state})
			for the system with $N=1$ particle of each species.
			After initialising it in the ground state
			the driving parameter $2\mu/\omega$ was linearly
			increased in the time $\tau$ until reaching the first
			zero of $\J(2\mu/\omega)$
			with $(2\mu/\omega)_\ab{max} \simeq 2.405$;
			the interaction parameter was $u=\kab/\Omega = 0.5$.
			The driving frequency used is $\omega/\Omega = 10$ (upper curve)
			and $\omega/\Omega = 1.2$ (lower curve).
			The increase with the higher frequency is in almost perfect agreement with
			the ideal curve	indicated by the dashed line.
			For the lower frequency the increase does not work and the
			\emph{Bell state} is not reached.
			}
	\label{N2_entanglement}
\end{figure}

	To quantify the entanglement of a state  $|\Psi \rangle $ we use the fidelity,
	i.e.\ the probability of the system to be in the
	\emph{Bell state} given by the	square of the state's overlap with
	$\psi_{\rm Bell}$:
	\begin{equation}
		P_\ab{Bell}=|\langle \psi_\ab{Bell} | \Psi \rangle |^2\;.
		\label{fidelity}
	\end{equation}
	Because our goal are fidelities close to one, more sophisticated
	entanglement measures
	(see e.g.\ refs.~\cite{Yukalov2003,Nielsen_Chuang}),
	which identify much less entangled states, are not necessary.
	As can be seen in fig.~\ref{N2_entanglement} entanglement is maximized
	by the procedure described above when
	the amplitude is changed adiabatically (for a value of say $\tau_\ab{max} = 2$
	the increase would be too fast and the maximal entanglement
	is not reached). On the other
	hand the driving frequency must not be too low. For $\omega/\Omega = 100$
	everything works excellently as expected; even for $\omega/\Omega = 10$
	we get very good results,
	but with $\omega/\Omega$ on the order of one a monotonic
	increase of $P_\ab{Bell}$ is not observed.

\section{Analytical calculation for the Bell state}

	In the spirit of refs.~\cite{CiracLewensteinMolmer1998,HolthausStenholm2001}
	we use the mean-field equations, in order to extend
	the two-particle dynamics to many Bosons,
	and construct a symmetrised superposition of two mean-field states.
	Within the mean-field approximation the system is reduced
	to four amplitudes describing the dynamics of condensate A (B),
	such that $|a_1|^2 = 1-|a_2|^2$ ($|b_1|^2 = 1-|b_2|^2$)
	is the expected fraction of condensate A (B) in the first well.
	The equations of motion for two condensates with equal number of particles
	($N_\ab{A} = N_\ab{B} = N$)
	are given by
	\begin{eqnarray}
		i \dot{a_1} &=& -\frac{1}{2}a_2  + 2N\frac{\ka}{\Omega} |a_1|^2 a_1
			+	4N \frac{\kab}{\Omega}|b_1|^2a_1 \label{mf_eqm_1}\\
		i \dot{a_2} &=& -\frac{1}{2}a_1  + 2N\frac{\ka}{\Omega} |a_2|^2 a_2
			+	4N \frac{\kab}{\Omega}|b_2|^2a_2 \\
		i \dot{b_1} &=& -\frac{1}{2}b_2  + 2N\frac{\kb}{\Omega} |b_1|^2 b_1
			+	4N \frac{\kab}{\Omega}|a_1|^2b_1 \\
		i \dot{b_2} &=& -\frac{1}{2}b_1  + 2N\frac{\kb}{\Omega} |b_2|^2 b_2
			+	4N \frac{\kab}{\Omega}|a_2|^2b_2\;, \label{mf_eqm_4}
	\end{eqnarray}
	where the dot means the derivative with respect to
	$\tau = \Omega \cdot t$ and $\ka$, $\kb$
	and $\kab$ are the interaction parameters given in
	eq.~(\ref{N-Hamiltonian}). In
	the following the intra-condensate interactions are equalised
	and the abbreviations $\alpha \equiv 2\frac{N\kappa}{\Omega}$
	and $\beta \equiv 2\frac{N\kab}{\Omega}$ are used.
	Experimentally, a similar situation could be achieved by choosing
	the numbers of particles in the condensates, such that
	$N_\ab{A} \ka = N_\ab{B} \kb $. The inter-condensate interaction
	parameter $\beta$ can be modulated by a
	Feshbach-resonance~\cite{
Feshbachresonanzen,MarteEtAl02}.

	To find stationary state solutions we apply the ansatz
	\begin{equation}
	  a_{1,2} = \psi_{1,2} \exp(-i\nu t) \;\; \text{and} \;\;
      b_{1,2} = \phi_{1,2} \exp(-i\mu t)\;.
	\end{equation}
	One finds two kinds of solutions, a balanced one, where the condensates are
	equally distributed on the two wells
	($\psi_1^2 = \psi_2^2 = \phi_1^2 = \phi_2^2 = \frac{1}{2}$)
	and an unbalanced solution, where condensate A/B is mainly in the
	first/second well. For $(2\beta-\alpha)> 0$ the amplitudes in this case are
	\begin{equation}
 		\label{MF_asym_1}
	 		\psi_{1/2} =  \phi_{2/1}
			=\frac{1}{\sqrt{2}}\left(1 \pm \sqrt{1-\frac{1}{(2\beta-\alpha)^2}}
			\right)^{1/2} \;.
	\end{equation}
	A second unbalanced solution is obtained by interchanging the indices 1 and 2.
 	If $(2\beta-\alpha)$ is negative, $\psi_1$ and $\psi_2$ in equation (\ref{MF_asym_1})
 	have opposite signs. As the resulting state is not the ground state in the limit
	of large interactions necessary for entanglement generation, this solution
	can be discarded here. The mean-field Hamiltonian corresponding to
	eqs.~(\ref{mf_eqm_1})-(\ref{mf_eqm_4}) reads:
	\begin{eqnarray}
		H  &=&\frac{\hbar \Omega}{2}\left[-(a_1 a_2^* + a_1^* a_2)+
		\alpha (|a_1|^4 + |a_2|^4) \right]
		\nonumber \\
		 &+&\frac{\hbar \Omega}{2} \left[-(b_1 b_2^* + b_1^* b_2)
		 +\alpha (|b_1|^4 + |b_2|^4)  \right]
		\nonumber \\
		&+&\hbar \Omega\, \beta \left( |a_1|^2 |b_1|^2 + |a_2|^2 |b_2|^2\right)
		\;.
	\end{eqnarray}
	The energies for the balanced and unbalanced states per particle therefore are given by
	\begin{equation}
		\frac{E_\ab{bal}}{N \hbar \Omega} =\frac{\alpha+\beta}{2} \mp 1
		\;,\quad
		\frac{E_\ab{ubal}}{N \hbar \Omega}
		=\alpha +\frac{\alpha - 3\beta}{2(2\beta-\alpha)^2}
		\;.
	\end{equation}
	When is the unbalanced state the ground state? It is, if
	 $E_\ab{ubal} < E_\ab{bal}$, leading to
	\begin{equation}
		\alpha - \beta + 2 + \frac{\alpha - 3\beta}{(2\beta-\alpha)^2} < 0\;.
		\label{mf_ineq}
	\end{equation}
	When again by driving the double-well the effective interactions
	are strongly increased (\ref{ueff}) only the difference of the intra- and inter-condensate
	interactions $\alpha$ and $\beta$ determine whether the balanced or the unbalanced
	state is the ground state.

	To compare the nonlinear mean-field dynamics to the $N$-particle dynamics, we
	need the $N$-particle counterparts of the mean-field states. By using the
	atomic coherent states \cite{HolthausStenholm2001}
	\begin{eqnarray}
		| \vartheta, \varphi \rangle_\ab{A} &=&
		\sum_{n_\ab{A}=0}^N \genfrac{(}{)}{0pt}{}{N}{n_\ab{A}}^{1/2}
		\cos^{n_\ab{A}}(\vartheta/2)
		 \sin^{N-n_\ab{A}}(\vartheta/2)
		 \nonumber\\
		 &\times& e^{i(N-n_\ab{A})\varphi}| n_\ab{A}, N-n_\ab{A} \rangle_\ab{A}
	\end{eqnarray}
	for condensate A (and B analogously)
	with $\cos(\vartheta/2) = \psi_1$,
	$\sin(\vartheta/2) = \psi_2$ and phase $\varphi = 0$
	we can construct a symmetrised superposition of two unbalanced states
	\begin{equation}
		|\Psi \rangle = \mathcal{N} \left(
		|\vartheta, 0 \rangle_{\rm A} | \vartheta, 0 \rangle_{\rm B}
		+ \overline{|\vartheta, 0 \rangle}_{\rm A} \overline{| \vartheta, 0 \rangle}_{\rm B}
		\right)\;,
	\end{equation}
	where the overlined term is obtained by replacing
	\mbox{$|n, N- n \rangle_\ab{A/B} $} by $|N- n, n \rangle_\ab{A/B}$.
	The probability $P_\ab{Bell}$ is then given by
	\begin{equation}
     P_\ab{Bell} = 2\mathcal{N}^2\left( \psi_1^{2N}+\psi_2^{2N} \right)^2\;.
	\end{equation}
	For large $N$
	the above expression is approximated by
	\begin{equation}
     P_\ab{Bell} \simeq \exp\left(-\frac {N\J\left(\frac{2\mu}{\omega}\right)^2}{2(2\beta-\alpha)^2} \right)
	 \label{mf_approx}\;.
	\end{equation}
\begin{figure}
	\subfigure{
	\includegraphics[scale=0.28,angle = -90]{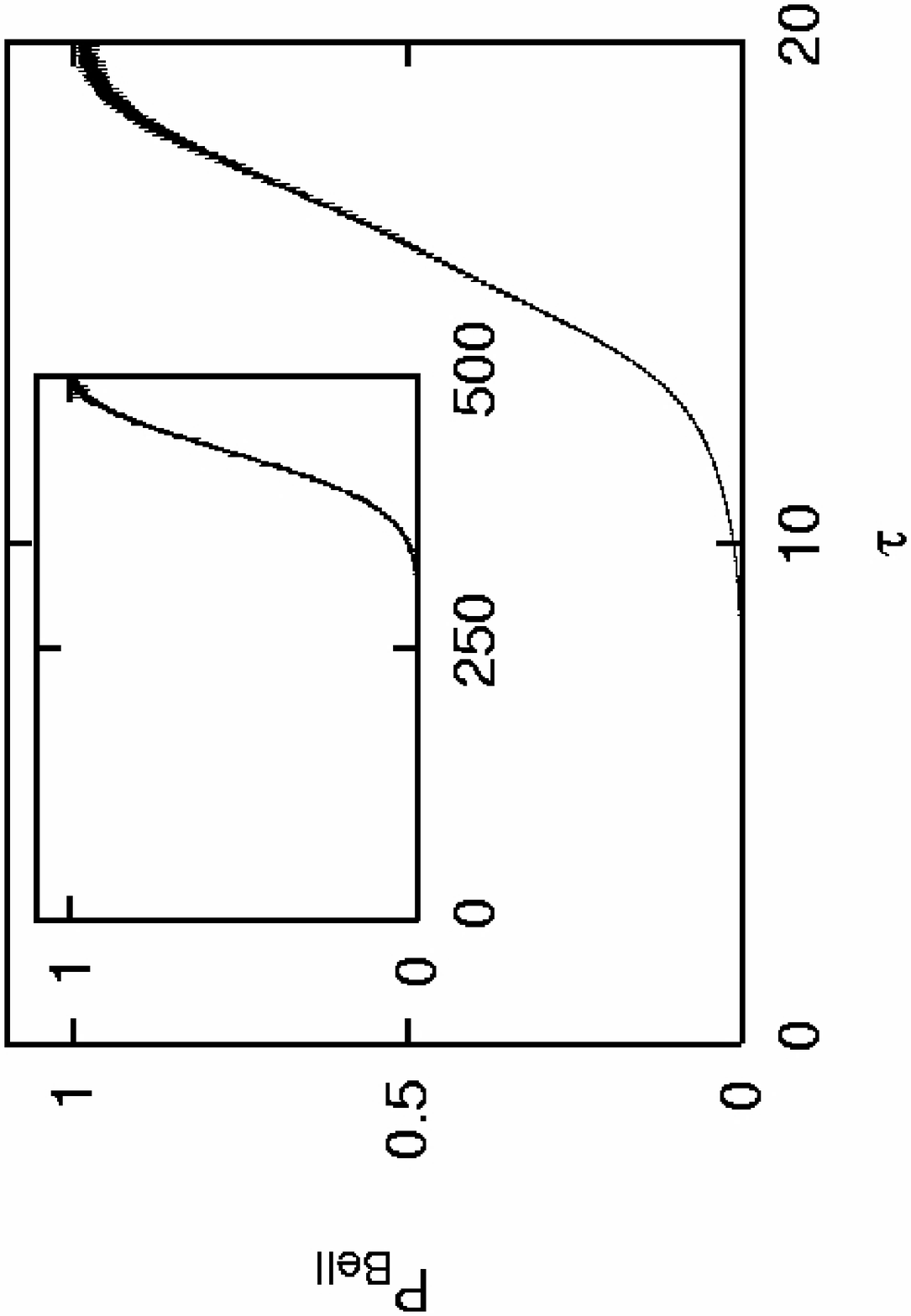}}
	\subfigure{
	\includegraphics[scale=0.28,angle = -90]{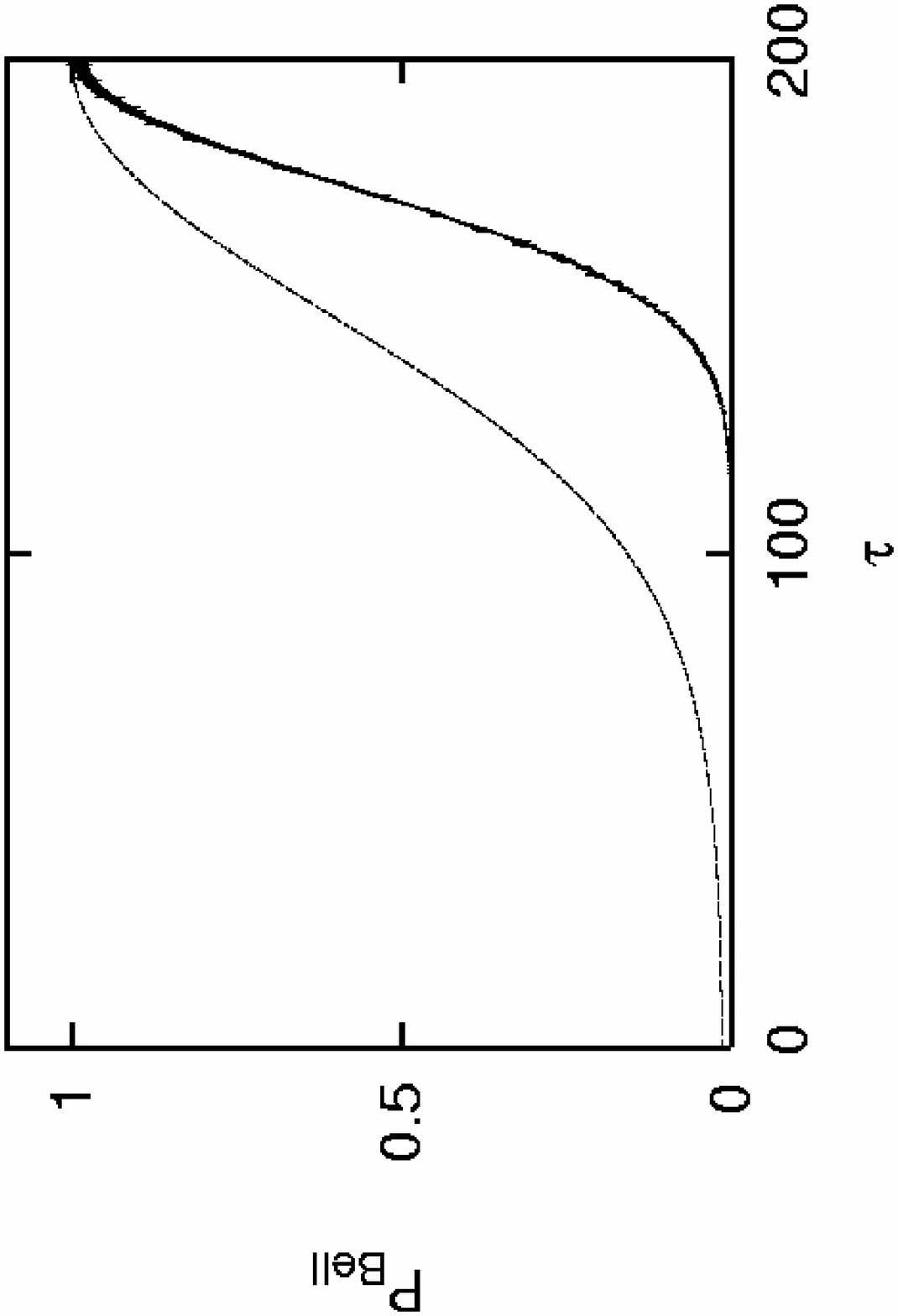}}
	\caption{Fidelity (\ref{fidelity}) for the system with $N=50$ particles of each species
		(cf.\ fig.~\ref{N2_entanglement}).
		After initialising it in the ground state
		the driving parameter $2\mu/\omega$ was linearly increased
		until reaching the first
		zero of $\J(2\mu/\omega)$
	; the used driving frequency
		was $\omega/\Omega = 40$.
		The upper plot shows that generating the \emph{Bell state} is possible on a short time scale
		with a fidelity higher than 96\un{\%};
		the interaction parameters were $\ka/\Omega=\kb/\Omega = 0.1$ and $\kab/\Omega = 0.115$.
		Inset: If the interaction parameters are chosen such that the initial state
		is not bimodal (see fig.~\ref{bell3d})
		with $\ka/\Omega=\kb/\Omega = 0.1$ and $\kab/\Omega = 0.11$, it takes longer to adiabatically
		reach the \emph{Bell state}. The fidelity is again very high ($>$ 97.5\un{\%}).
		Lower figure: although the unbalanced state is already quite entangled
		when it becomes the ground state for $\tau \approx 80$
		(cf.\ eq.~(\ref{mf_ineq})), the analytic curve indicated by the dashed line
		(cf.\ eq.~(\ref{mf_approx}))
		gives a good qualitative description of this process.
		The used interaction parameters here
		are $\ka/\Omega=\kb/\Omega = 0.0016$ and $\kab/\Omega = 0.0133$;
		the fidelity is $>$ 97.5\un{\%}.
	\label{N-particle}}
\end{figure}

\section{$N$-particle dynamics and entanglement generation}

\begin{figure}
  \subfigure{
  	\includegraphics[scale = 0.6]{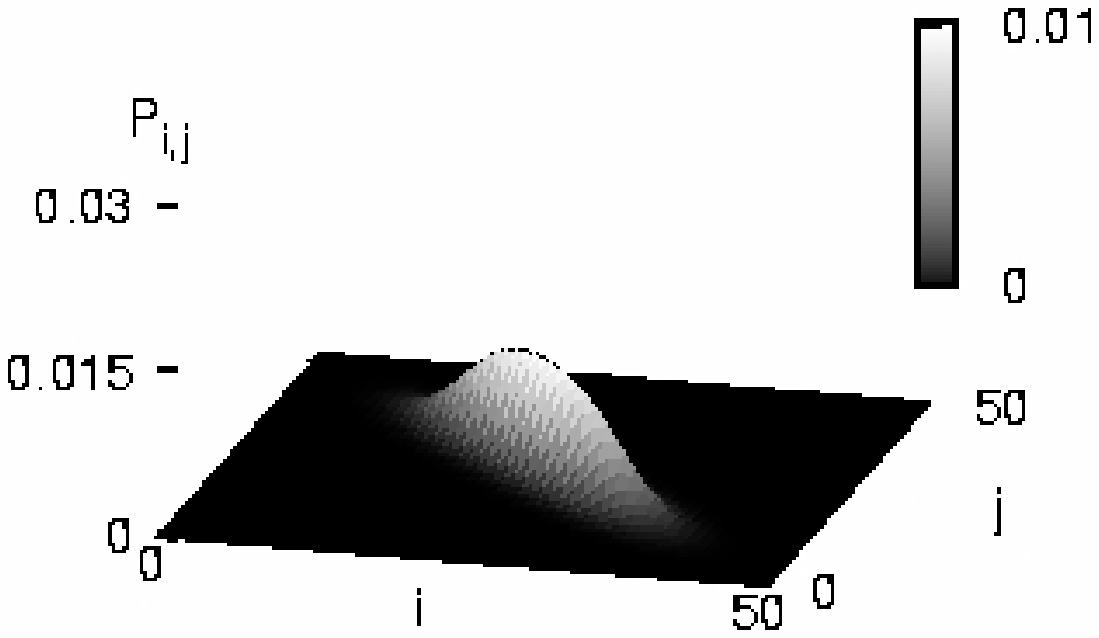}}
	\subfigure{
  	\includegraphics[scale = 0.6]{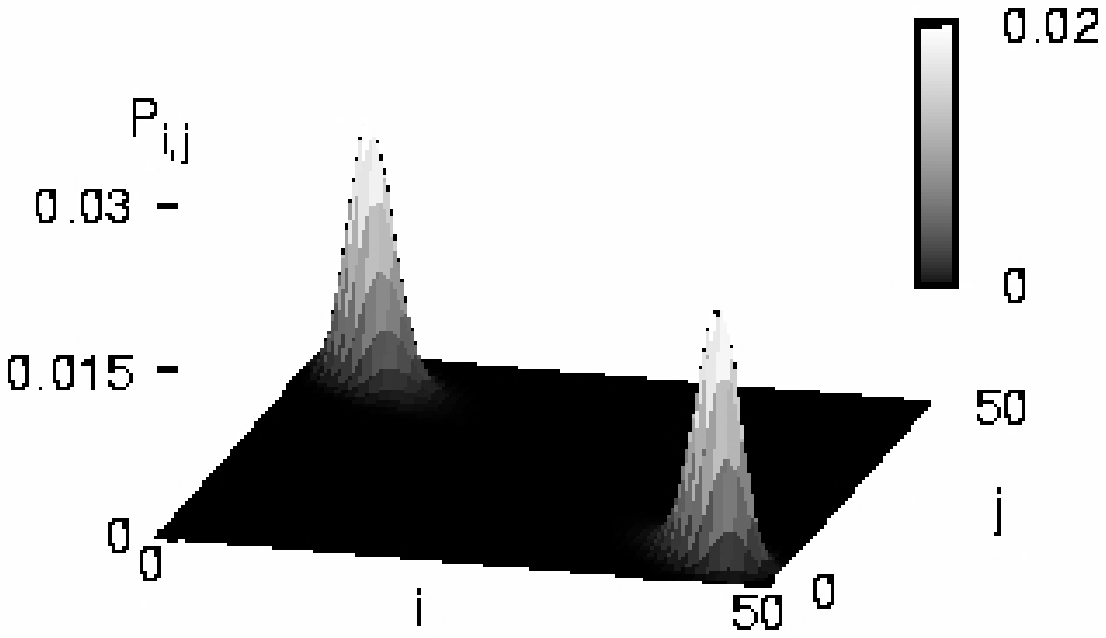}}
  \caption{Probabilities of the system to be in the Fock state
  	$|i,j \rangle $, where $i$ and $j$ are the Fock indices. The system was initialised in the
	ground state, which is in this case
	(same parameters as in the inset of fig.~\ref{N-particle})
	a quite balanced state (upper figure: system at $\tau=0$),
	and then driven with linearly increasing driving parameter $2\mu/\omega$.
	With increasing time the system's state becomes more and more unbalanced
	(lower figure: system at $\tau = 250$) until it reaches the Bell state  at
	$\tau_\ab{max}=500$ (cf.\ inset of fig.~\ref{N-particle}).
	}
  \label{bell3d}
\end{figure}

	To demonstrate entanglement generation numerically,
    we use 100 particles (50 of each kind), which is close to the
 	experimentally reachable domain \cite{ChuuSchreckMeyrath2005}.
	In experiments the decision for small condensates makes sense, because
	decoherence times induced by particle loss are long.
	Again we prepare the system in the ground state and drive it with
	linearly in time $\tau$ increasing driving
	parameter $2\mu/\omega$ up to the first zero of $\J$.
	Figure~\ref{N-particle} shows the emergence of the highly entangled
	state under the influence of a periodic force-field.
	The parameters are chosen depending on the desired behaviour:
	generation of entanglement on very short time scales
	(fig.~\ref{N-particle}, top), good agreement with the
	analytical calculation (eq.~(\ref{mf_approx}))
	or generation of entanglement starting with a balanced initial state
	(see fig.~\ref{bell3d}). Although the mean-field equations are
	valid in the limit $N\to\infty$ and $\kappa\to 0$ such that
	$N\kappa$ is constant, the predictions obtained by our calculations
	 qualitatively agree with the numerical
	results as long as $(2\beta-\alpha)$ is not too high in the regime,
	where the balanced state is the ground state.

	In fig.~\ref{bell3d} we show the occupation of states by plotting
	the Fock occupation numbers $i$ and $j$ and the corresponding probabilities
	$| \langle \Psi |i,j \rangle|^2$.
	The probabilities move to the outer (more unbalanced) states
	while increasing the parameter $2\mu/\omega$
	and at the end only the two states $| N, 0 \rangle$ and $| 0, N \rangle$
	are nearly equally occupied.
	So a quite balanced state is changed into a totally unbalanced
	and maximally entangled state.
\begin{figure}
   	\includegraphics[angle=-90,scale = 0.3]{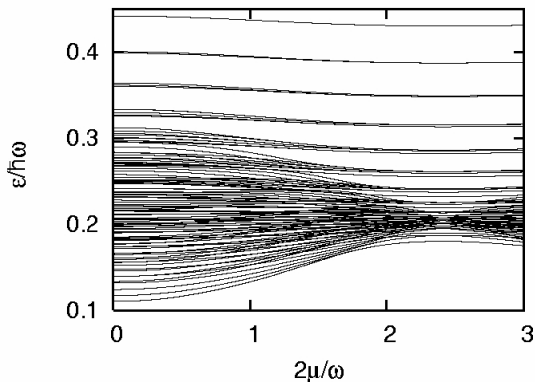}
	\caption{Part of the quasienergy spectrum for the system~(\ref{N-Hamiltonian}) with $N=10$ particles
	of each kind and driving frequency $\omega/\Omega = 100$,  $\ka/\Omega = \kb/\Omega = 0.1$ and 
	$\kb/\Omega = 0.125$.  At $2\mu/\omega\simeq 2.4$  the ground-state,
	the lowest line, is quasi-degenerate (as many other states, cf.\ ref.~\cite{EckardtWeissHolthaus2005}).
	At this point $\J(2\mu/\omega)$ is nearly zero and therefore tunnelling is strongly suppressed. The coherent 
        control used in this letter involves adiabatic following of the ground state. While transitions to the 
	state which follows the first excited state are suppressed for symmetry reasons (cf.\ ref.~\cite{Haroche}), 
	the energy difference to the higher states gives a measure for the timescale on which the amplitude can be 
	changed. (For $N=50$ most of the spectrum would become dense. However, for high driving frequencies the lowest two 
	quasienergies are still well separated from the other quasienergies -- otherwise the high fidelities of 
	fig.~\ref{N-particle} would not have been reached).
}
  \label{quasienergies}
\end{figure}

	Within the validity of the
	two-mode approximation (cf.\ ref.~\cite{Jinasundera_Weiss}),
	the fidelity (\ref{fidelity}) can be raised by using higher
	driving frequencies $\omega/\Omega$. As the entanglement generation involves adiabatic following 
        of Floquet states (see ref.~\cite{EckardtWeissHolthaus2005} and references therein), the time-scales 
        on which the amplitude will have to be changed in future experiments can be deduced from calculating 
        the quasienergies (fig.~\ref{quasienergies}). For higher particle numbers than chosen in 
	fig.~\ref{quasienergies} and not too high driving frequencies, avoided crossings occur
	which are not important for entanglement generation 
        as long as the driving frequency still is in the high frequency limit 
        (fig.~\ref{N-particle}; see also ref.~\cite{EckardtWeissHolthaus2005}). To find the best 
	parameters for a given experiment with respect
	to low decoherence and large entanglement and to optimise the protocol~$\mu(\tau)$, 
        i.e., the way the driving amplitude is changed in time, one might
	employ the methods of optimal control theory~\cite{AssionBaumert1998}.

\section{Conclusion}

	We demonstrated that the generation of a highly
	entangled state with a binary condensate in
	a double-well potential	is possible.
	By driving the double-well
	with an increasing amplitude the ground state is turned into
	a \emph{Bell state} provided that the inter-condensate interaction
	is first slightly increased via a Feshbach-resonance.
	The fidelities of our results within the model of eq.~(\ref{N-Hamiltonian})
	are even on short time scales above $96\un{\%}$ which is considerably higher than
        the fidelity obtained in similar models for single-species condensates~\cite{MicheluJakschCirac2003,MahmudPerryReihardt1998,Jinasundera_Weiss}.

\acknowledgments
We would like to thank M.~Holthaus and Y.~Castin for valuable comments.
N.~T.\ acknowledges support by the Studienstiftung des deutschen Volkes; C.~W.\ acknowledges funding by the German Research Foundation (DFG) through Grant WE 2867/2-1 and the European Union through contract MEIF-CT-2006-038407.


\begin{thebibliography}{0}
\bibitem{Feshbachresonanzen}
\title{One possible realisation of such a binary condensate could be two hyperfine-states of $^{87}$Rb, for which Feshbach-resonances between different hyperfine-states have been observed experimentally~\cite{MarteEtAl02}.}
\end{thebibliography}
\end{document}